\def\thetabar{\overline{\theta}}
\def\gagg{g_{a\gamma\gamma}}
\def\ma{m_a}
\def\meV{$\mu$eV}

\documentclass{iopart}
\usepackage{amssymb}
\usepackage{graphicx}
\begin{document}

\title{Axions as Dark Matter Particles}

\author{Leanne D.\ Duffy$^1$ and Karl van Bibber$^2$}
\address{$^1$ Los Alamos National Laboratory, Los Alamos, NM 87545, USA}
\address{$^2$ Lawrence Livermore National Laboratory, Livermore, CA 94550, USA and the Naval Postgraduate School, Monterey, CA 93943, USA}
\ead{lduffy@lanl.gov, kvanbibber@llnl.gov}

\begin{abstract}
We review the current status of axions as dark matter.  Motivation,
models, constraints and experimental searches are outlined.  The axion remains
an excellent candidate for the dark matter and future experiments, particularly
the Axion Dark Matter eXperiment (ADMX), will cover a large fraction of the
axion parameter space.
\end{abstract}

\maketitle

\section{Introduction}

Despite our knowledge of dark matter's properties, what it consists of is
still a mystery.
The standard
model of particle physics does not
contain a particle that qualifies as dark matter.
Extensions to the standard model do, however, provide viable particle 
candidates.  The axion, the pseudo-Nambu-Goldstone boson of the
Peccei-Quinn solution to the strong CP problem
\cite{Peccei:1977hh,Peccei:1977ur,Weinberg:1977ma,Wilczek:1977pj}, is a
strongly motivated particle candidate.  We review the strong CP problem and its resulting
axion in section~\ref{sec:strongCP}.
  
In the early universe, cold
axion populations arise from vacuum realignment
\cite{Abbott:1982af,Preskill:1982cy,Dine:1982ah}
 and string and wall decay 
\cite{Davis:1985pt,Davis:1986xc,Harari:1987ht,Vilenkin:1986ku,Davis:1989nj,Dabholkar:1989ju,Battye:1993jv,Battye:1994au,Yamaguchi:1998gx,Hagmann:2000ja,Chang:1998tb,Lyth:1991bb,Nagasawa:1994qu}.  Which mechanisms
contribute depends on whether the Peccei-Quinn symmetry breaks
before or after inflation.  These cold axions were never in thermal
equilibrium with the rest of the universe and could provide the missing dark matter.  
The cosmological production of cold axions is outlined in section~\ref{sec:cosmo}.

Current constraints on the axion parameter space, from astrophysics, cosmology
and experiments, are reviewed in 
section~\ref{sec:bounds}.
The signal observed in direct detection experiments depends on the phase-space distribution of dark 
matter axions.  In section~\ref{sec:phasespace}, we discuss possible structure for dark matter in our galactic halo and
touch on the implications for detection.  Experimental searches and their current
status are discussed in section~\ref{sec:exp}.  Finally, the current status of the QCD
axion is summarized in section~\ref{sec:sum}.

This is a brief review, designed to overview the current status of axions as dark
matter.  For more advanced details, we refer the reader to the extensive literature (e.g. reviews can be found in refs.~\cite{Kim:1986ax,Cheng:1987gp,Turner:1989vc,Raffelt:1990yz,Kim:2008hd}).  

\section{The strong CP problem and the axion}
\label{sec:strongCP}

\subsection{Strong CP problem}
\label{ssec:strongcp}
The strong CP problem arises from the non-Abelian nature of the QCD gauge symmetry, or colour 
symmetry.  Non-Abelian gauge potentials have disjoint sectors that cannot be 
transformed continuously into one another.  Each of these vacuum configurations can be labelled
by an integer, the topological winding number, $n$.  Quantum tunnelling occurs between
vacua.  Consequently, the gauge invariant QCD vacuum state is a superposition of these states,
i.e.,
\begin{equation}
|\theta\rangle=\sum_{n}e^{-in\theta}|n\rangle \; .
\label{eq:QCDvac}
\end{equation}
The angle, $\theta$, is a parameter which describes the QCD vacuum state, $|\theta\rangle$.

In the massless quark limit, QCD possesses a classical chiral symmetry.
However, this symmetry is not present in the full quantum theory due to the Adler-Bell-Jackiw
anomaly \cite{Adler:1969gk,Bell:1969ts}.
In the full quantum theory, including quark masses, the
physics of QCD remains unchanged under the following transformations of the quark fields,
$q_i$, quark masses, $m_i$, and vacuum parameter, $\theta$:
\begin{eqnarray}
q_{i}&\rightarrow& e^{i\alpha_i\gamma_{5}/2}q_{i}\label{eq:qtrans}\\
m_{i}&\rightarrow& e^{-i\alpha_i}m_{i}\label{eq:mtrans}\\
\theta&\rightarrow&\theta-\sum_{i=1}^N\alpha_i\label{eq:thetatrans} \; ,
\end{eqnarray}
where the $\alpha_i$ are phases and $\gamma_5$ is the usual product of gamma matrices.
This is not a symmetry of QCD due to the change in $\theta$.

The transformations of
Eq.~(\ref{eq:qtrans}) through Eq.~(\ref{eq:thetatrans}) can be used to move 
phases between the quark masses and $\theta$.  However, the
quantity, 
\begin{equation}
\thetabar\equiv\theta-\arg\det\mathcal{M}=\theta-\arg(m_1m_2...m_N) \;,
\end{equation}
where $\mathcal{M}$ is the quark mass matrix, is invariant and thus observable, unlike $\theta$.  

The presence of $\theta$ in QCD violates the discrete symmetries P and CP.
However, CP violation has not been observed in QCD.
An electric dipole moment for the neutron is the most easily observed
consequence of QCD, or strong, CP violation.  $\theta$ results in a neutron
electric dipole moment of
\cite{Kim:1986ax,Cheng:1987gp,Turner:1989vc,Raffelt:1990yz},
\begin{equation}
|d_n|\sim10^{-16}\;\thetabar\; e\; \mathrm{cm} \, ,
\label{eq:NEDMtheory}
\end{equation}
where $e$ is the electric charge.
The current experimental limit is \cite{Harris:1999jx}
\begin{equation}
|d_n|<6.3\times10^{-26}\; e \;\mathrm{cm} \; 
\label{eq:NEDMexp}
\end{equation}
and thus, $|\thetabar|\lesssim 10^{-9}$.  
There is no natural reason to expect $\thetabar$ to be this small.
CP violation occurs in the standard model by allowing the quark
masses to be complex and thus the natural value of 
$\theta$ is 
expected to be of order one.  This
is the strong
CP problem, i.e.\ the question of why the angle $\thetabar$ should be nearly
zero, despite the presence of CP violation in the standard model.

The Peccei-Quinn (PQ)
solution \cite{Peccei:1977hh,Peccei:1977ur} to this problem results in 
an axion \cite{Weinberg:1977ma,Wilczek:1977pj}.
While other solutions to the strong CP problem have been proposed, the presence of the
axion in the PQ solution makes it the most interesting when searching for the dark
matter of the universe.  Thus, we focus on this solution only in the following. 

\subsection{The Peccei-Quinn solution}

The axion is the pseudo-Nambu-Goldstone boson from the
Peccei-Quinn solution to the strong CP problem
\cite{Peccei:1977hh,Peccei:1977ur,Weinberg:1977ma,Wilczek:1977pj}.
In the PQ solution, 
$\thetabar$ is promoted from a parameter to
a dynamical variable.  This variable relaxes to the minimum of its potential and hence is small. 

To implement the PQ mechanism, a global
symmetry, $U(1)_{PQ}$, is introduced.  This symmetry possesses a colour anomaly and
is spontaneously broken.
The axion is the resulting Nambu-Goldstone boson and its field,
$a$, can be redefined to absorb the parameter $\thetabar$.  While initially massless, 
non-perturbative 
effects, which make QCD $\thetabar$ dependent, also result in a potential for the axion.
This potential causes the axion to
acquire a mass and relax to
the CP conserving minimum, solving the strong CP problem.  

As there are no degrees of freedom available for the axion in the standard model, new fields 
must be added to realize the PQ solution.  In the original, 
Peccei-Quinn-Weinberg-Wilczek (PQWW) axion model, an extra Higgs doublet
was used.  We review this model to demonstrate the PQ
mechanism. 

Assume that one of the two Higgs doublets in the model, $\phi_{u}$,
couples
to up-type quarks with strengths $y_i^u$ and the other, $\phi_{d}$, couples to 
down-type quarks with strengths $y_i^d$, where $i$ gives the variety of up- or down-type
quark. 
We label the
up- and down-type quarks $u_{i}$ and $d_{i}$, respectively
(rather than
$q_{i}$, as in the previous section).  With a total of $N$ quarks, there are $N/2$
up-type quarks and down-type quarks.  The
leptons may
acquire mass via Yukawa couplings to either of the Higgs doublets or to a
third
Higgs doublet.  We ignore this complication here and simply
examine the couplings to quarks.  

The quarks acquire their masses from the
expectation values of the neutral components of the Higgs, $\phi_{u}^{0}$ and
$\phi_{d}^{0}$.  The mass generating couplings are
\begin{equation}
\mathcal{L}_{m}=
y^{u}_{i}u^{\dagger}_{Li}\phi^{0}_{u}u_{Ri}
 +y^{d}_{i}{d}^{\dagger}_{Li}\phi^{0}_{d}d_{Ri}+\textrm{h.c.} \; .
\label{eq:Lmass}
\end{equation}
Peccei and Quinn chose the Higgs potential to be
\begin{equation}
V(\phi_{u},\phi_{d})=-\mu_{u}^{2}\phi_{u}^{\dagger}\phi_{u}
  -\mu_{d}^{2}\phi_{d}^{\dagger}\phi_{d}
  +\sum_{i,j}a_{ij}\phi_{i}^{\dagger}\phi_{i}\phi_{j}^{\dagger}\phi_{j}
  +\sum_{i,j}b_{ij}\phi_{i}^{\dagger}{\phi}_{j}
    {\phi}_{j}^{\dagger}\phi_{i} \; ,
\end{equation}
where the matrices $(a_{ij})$ and $(b_{ij})$ are real and symmetric and the
sum is over the two types of Higgs fields. 
With
this choice of potential, the full Lagrangian
has a global $U_{PQ}(1)$ invariance,
\begin{eqnarray}
\phi_{u}&\rightarrow& e^{i2\alpha_{u}}\phi_{u}\\
\phi_{d}&\rightarrow& e^{i2\alpha_{d}}\phi_{d}\\
u_{i}&\rightarrow& e^{-i\alpha_{u}\gamma_{5}}u_{i}\label{eq:utrans}\\
d_{i}&\rightarrow& e^{ -i \alpha_{d}\gamma_{5}}d_{i}\label{eq:dtrans}\\
\thetabar&\rightarrow&\thetabar-N(\alpha_{u}+\alpha_{d})\label{eq:thetatrans2}\; .
\end{eqnarray}

When the electroweak symmetry breaks, the neutral Higgs components acquire
vacuum expectation values:
\begin{eqnarray}
\langle\phi^{0}_{u}\rangle=v_{u}e^{iP_{u}/v_{u}} \label{eq:phiu}\\
\langle\phi^{0}_{d}\rangle=v_{d}e^{iP_{d}/v_{d}} \label{eq:phid}\; .
\end{eqnarray}
One linear combination of the Nambu-Goldstone fields, $P_{u}$ and $P_{d}$, 
is the longitudinal component of the Z-boson,
\begin{equation}
h=\cos \beta_{v}P_{u}-\sin \beta_{v} P_{d} \; .
\label{eq:h}
\end{equation}
The orthogonal combination is the axion field,
\begin{equation}
a=\sin \beta_{v}P_{u}+\cos \beta_{v} P_{d} \; .
\label{eq:a}
\end{equation}
Using Eqs.~(\ref{eq:phiu}), (\ref{eq:phid}), (\ref{eq:h})
and
(\ref{eq:a}) with Eq.~(\ref{eq:Lmass}), the axion couplings to quarks arise
from
\begin{equation}
-\mathcal{L}_{m}=m^{u}_{i}u^{\dagger}_{Li}e^{i\frac{\sin\beta_{v}}{v_{u}}a}u_{Ri}
 +m^{d}_{i}{d}^{\dagger}_{Li}e^{i\frac{\cos\beta_{v}}{v_{d}}a}d_{Ri}+\textrm{h.c.}
\; ,
\end{equation}
where $m^{u}_{i}=y^{u_{i}}v_{u}$ and $m^{d}_{i}=y^{d}_{i}v_{d}$.  The axion
field dependence can be moved from the mass terms using the transformations of
Eqs.~(\ref{eq:utrans}), (\ref{eq:dtrans}) and (\ref{eq:thetatrans2}). 
Direct couplings between the axion and quarks will remain in the
Lagrangian, through the quark kinetic term.  Defining $v=\sqrt{v_{u}^{2}+v_{d}^{2}}$,
the
corresponding change in $\thetabar$ is
\begin{equation}
\thetabar\rightarrow\thetabar-N(v_{u}/v_{d}+v_{d}/v_{u})a/v \; ,
\label{eq:aeattheta}
\end{equation}
which can be absorbed by a redefinition of the axion field.  

Non-perturbative QCD effects explicitly break the
PQ symmetry, but do not become important until confinement occurs.
These effects give the axion field a 
potential and when significant, the field relaxes to the CP conserving minimum.
  Hence the PQ mechanism, which replaces $\thetabar$ with
the dynamical axion field, solves the strong CP problem.

Under the PQWW scheme, the axion mass is tied to the electroweak
symmetry breaking scale, resulting in a mass of the order of 100 keV.
This heavy PQWW axion has been ruled out by observation, as
discussed in Section~\ref{ssec:labbounds}.  This does not, however, eliminate
the possibility of solving the strong CP problem with an axion.  In the 
following section,
we discuss viable axion models.

\subsection{Axion models}

``Invisible''
axion models, named so for their extremely weak couplings, are still
possible.
In an invisible axion model, the PQ symmetry is decoupled from the
electroweak scale and is spontaneously broken at a much higher
temperature, decreasing the axion mass and coupling strength.
Two benchmark, invisible axion models exist:  the
Kim-Shifman-Vainshtein-Zakharov (KSVZ) \cite{Kim:1979if,Shifman:1979if} and
Dine-Fischler-Srednicki-Zhitnitsky (DFSZ) \cite{Dine:1981rt,Zhitnitsky:1980tq} models.
In both these models, an
axion with permissable mass and couplings
results.

In the KSVZ model, the only
Higgs doublet is that of the standard model.  The axion is introduced as the
phase of an additional electroweak singlet scalar field.  The known
quarks cannot directly couple to such a field, as this would lead to
 unreasonably large quark masses.  Instead, the scalar
is coupled to an additional heavy quark, also an electroweak singlet.  The
axion couplings are then induced by the interactions of the heavy quark
with the other fields.

The DFSZ model has two Higgs doublets, as in the PQWW model, and an
additional electroweak singlet scalar.  It is the electroweak singlet
which acquires a vacuum expectation value at the PQ symmetry breaking scale.  The scalar does
not couple directly to quarks and leptons, but via its interactions with the
two Higgs doublets.

PQ symmetries also occur naturally in string theory, via string compactifications, 
and are always broken by 
some type of instanton.  While this could be expected to make the axion an outstanding dark
matter candidate, string models favour a value of the PQ scale that is much higher than that allowed 
by cosmology (see the discussion in \ref{sec:cosmo}).  A review of the current situation
can be found in Ref.~\cite{Svrcek:2006yi}.  As discussed in \cite{Svrcek:2006yi}, it is difficult to push the PQ scale
far below $1.1\times10^{16}$~GeV and easier to instead increase its value.  

Generically, axion couplings to other particles are inversely proportional to $f_a$, however, the exact
strength of these couplings are model dependent.  For example, the coupling between axions and photons can be written,
\begin{equation}
\mathcal{L}_{a\gamma\gamma}   = g_{a\gamma\gamma}a\mathbf{E}\cdot\mathbf{B} \; ,
\label{eq:aEB}
\end{equation}
where $\mathbf{E}$ and $\mathbf{B}$ are the electromagnetic field components.  The coupling constant is
\begin{equation}
g_{a\gamma\gamma}=\frac{g_{\gamma}\alpha}{\pi f_a} \; ,
\end{equation}
where $\alpha$ is the electromagnetic fine structure constant, $10^9$~GeV~$\lesssim f_a \lesssim 10^{12}$~GeV 
is the axion decay constant, of the
order of the PQ scale, and $g_\gamma$ is a constant containing the model dependence.  Explicitly,
\begin{equation}
g_{\gamma}=\frac{1}{2}\left(\frac{E}{N}-\frac{2(4+z)}{3(1+z)}\right) \; ,
\end{equation}
where $z$ is the ratio of the up and down quark masses, $N$ is the axion colour anomaly and $E$, the axion
electromagnetic anomaly \cite{Rosenberg:2000wb}.  The term containing the ratios of light quark masses is 
approximately equal to $1.95$.  The model dependence arises through the ratio, $E/N$.  In grand-unifiable models,
$E$ and $N$ are related and $E/N=8/3$.  The
DFSZ axion model falls into this category and in this case, $g_{\gamma}=0.36$.  For a KSVZ axion, $E=0$ and
$g_{\gamma}=-0.97$.  

It is possible for an axion to solve the strong CP
problem, as shown by the existence of the KSVZ and DFSZ axion models.  
While significant for that alone, the axion also provides an
interesting candidate for the cold dark matter of the universe.

%

\section{Cosmological production of axions}
\label{sec:cosmo}

\subsection{Properties of axion dark matter}

Axions satisfy the two criteria necessary for cold dark matter:  (1) a
non-relativistic population of axions could be present in our universe in
sufficient quantities to provide the required dark matter energy density and
(2) they are effectively collisionless, i.e., the only significant long-range 
interactions
are gravitational. 

Despite having a very small mass \cite{Kim:1986ax,Cheng:1987gp,Turner:1989vc,Raffelt:1990yz},
\begin{equation}
m_a \simeq6\times10^{-6}\; \textrm{eV}\;\left(\frac{10^{12} \mathrm{GeV}}{f_a}\right)
\; ,
\label{eq:ma}
\end{equation}
axion dark matter is non-relativistic, as 
cold populations are produced out of equilibrium.
There are three mechanisms via which cold axions are
produced: vacuum realignment 
\cite{Abbott:1982af,Preskill:1982cy,Dine:1982ah}, string decay 
\cite{Davis:1985pt,Davis:1986xc,Harari:1987ht,Vilenkin:1986ku,Davis:1989nj,Dabholkar:1989ju,Battye:1993jv,Battye:1994au,Yamaguchi:1998gx,Hagmann:2000ja,Chang:1998tb}
and domain wall decay \cite{Chang:1998tb,Lyth:1991bb,Nagasawa:1994qu}.
In this section, we discuss the history of the axion field as the universe expands and cools to
see how and when axions are produced.  We also review vacuum realignment production 
in detail, as there will always be a contribution to the
cold axion populations from this mechanism and, as discussed below, it may provide the only contribution.
A complete description of the cold axion populations can be found in ref.~\cite{Sikivie:2006ni}.

\subsection{Topological axion production}

There are two important scales in dark matter axion production.  The
first is the temperature at which the PQ symmetry breaks, $T_{PQ}$.  Which
of the three mechanisms contribute significantly to the cold axion population
depends on whether this temperature is greater or
less than the inflationary reheating temperature,
$T_{R}$.  The second scale is the
temperature at which the axion mass, arising from non-perturbative
QCD effects, becomes significant.  
At high temperatures, the QCD effects are
not significant and the axion mass is negligible \cite{Gross:1980br}.  The
axion mass becomes important at a critical time, $t_1$, when 
$m_a t_1\sim1$ \cite{Abbott:1982af,Preskill:1982cy,Dine:1982ah}.  
The temperature of the universe at $t_1$ is $T_1\simeq1$ GeV.

The PQ symmetry is unbroken at early times and temperatures greater than
$T_{PQ}$.    
At $T_{PQ}$, 
it breaks spontaneously and the axion field, proportional to the 
phase of the complex 
scalar field acquiring a vacuum expectation value, may have any value. 
The phase varies continously, changing by order one from one horizon to the 
next.  Axion strings appear as topological defects.

If $T_{PQ}>T_R$, the axion field is homogenized over vast distances and the
string density is diluted by inflation, to a point
where it is extremely unlikely that our visible universe contains any axion strings.
In the case $T_{PQ}<T_R$, the axion field
is not homogenized and
strings radiate cold, massless axions until non-perturbative QCD 
effects become
significant at temperature, $T_1$.  Agreement has not been reached on the 
expected spectrum of axions from string radiation and there are two possibilities.
Either strings oscillate many times before they completely decay and axion production is strongly peaked around a dominant mode \cite{Davis:1985pt,Davis:1986xc,Vilenkin:1986ku,Davis:1989nj,Dabholkar:1989ju,Battye:1993jv,Battye:1994au,Yamaguchi:1998gx}
or much more rapid decay occurs, producing a spectrum inversely proportional to momentum \cite{Harari:1987ht,Hagmann:1990mj}.  Rapid decay produces $\sim70$ times less axions than slow
string decay, leading to different cosmological bounds on the axion mass (see
section~\ref{ssec:cosmobounds}).

When the universe cools to $T_1$,  
the axion strings become the boundaries of $N$ domain walls.  For $N=1$, 
the walls 
rapidly radiate cold
axions and decay (domain
wall decay).  
If $N>1$, the domain wall problem occurs \cite{Sikivie:1982qv}
because the vacuum is multiply degenerate and there is at least one domain
wall per horizon.  These walls will end up dominating the energy density 
and cause the universe to expand as $S\propto t^2$, where $S$ is the scale
factor.  Although other solutions
to the domain wall problem have been proposed \cite{Chang:1998tb}, we
assume here that $N=1$ or $T_{PQ}>T_{R}$.

Thus, if $T_{PQ}<T_R$, string and wall decay contribute
to the axion energy density.  If $T_R<T_{PQ}$, and the axion string
density is diluted by inflation, these mechanisms do not 
contribute significantly to the density of cold axions.  Then, only
vacuum realignment will contribute a significant amount. 

\subsection{Vacuum realignment mechanism}
\label{ssec:vacalign}

Cold axions will be produced by vacuum realignment, independent
of $T_R$.  Details of this method are discussed below, but the general
mechanism is as follows.  At $T_{PQ}$, the axion field amplitude may have any value.
If $T_{PQ}>T_R$, homogenization will occur due to inflation and the axion field will
be single valued over our visible universe. 
Non-perturbative QCD effects cause a potential for the axion field.  When
these effects become significant, the axion field will begin to
oscillate in its potential.  These oscillations do not decay and 
contribute to the local energy density as non-relativistic matter.  Thus, a
cold axion population results from vacuum realignment, regardless of the 
inflationary reheating temperature.

To illustrate vacuum realignment, consider a toy axion model
with one complex scalar field, $\phi(x)$, in addition to the standard model
fields.  The 
potential for $\phi(x)$ in our toy model is
\begin{equation}
V(\phi)=\frac{\lambda}{4}(|\phi|^2-v_a^2)^2 \; ,
\end{equation}
When the universe cools to
$T_{PQ}\sim v_a$, $\phi$ acquires a vacuum expectation value,
\begin{equation}
\langle \phi \rangle = v_a \exp(i\theta(x))\; .
\end{equation}
The relationship between the axion field, $a(x)$, and $\theta(x)$ is
\begin{equation}
a(x)\equiv v_a \theta(x)\; .
\label{eq:axang}
\end{equation}
The axion decay constant in this model is
\begin{equation}
f_a\equiv\frac{v_a}{N}\; .
\label{eq:PQscale}
\end{equation}
For the following discussion, we set $N=1$.

At $T\sim\Lambda$, where $\Lambda$ is the confinement scale,
non-perturbative QCD effects give the axion a mass.
An effective potential,
\begin{equation}
\widetilde{V}(\theta)=m_a^2(T) f_a^2(1-\cos \theta)\; ,
\end{equation}
is produced.
The axion acquires mass, $m_a$, due to the curvature of the potential at
the minimum.  This mass is temperature, and thus time, dependent due to the temperature
dependence of the potential \cite{Gross:1980br}.

In a Friedmann-Robertson-Walker universe, the equation of motion for $\theta$ is
\begin{equation}
\ddot{\theta}+3H(t)\dot{\theta}-\frac{1}{S^{2}(t)}\nabla^{2}\theta
+m_{a}^{2}(T(t))\sin(\theta)=0 \; ,
\end{equation}
where $S(t)$ is the scale factor and $H(t)$, the Hubble constant, at time $t$. 
Near $\theta=0$, $\sin(\theta)\simeq \theta$.

We now restrict the discussion to the zero momentum mode.  This is the only
mode with significant occupation when $T_{PQ}>T_{R}$, so the final energy
density calculated will be for this case.  When $T_R>T_{PQ}$, 
higher modes will also be occupied.  
For the zero momentum mode,
neglecting spatial derivatives, the equation of motion reduces to
\begin{equation}
\ddot{\theta}+3H(t)\dot{\theta}+m_{a}^{2}(t)\theta=0 \; ,
\end{equation}
i.e., the field satisfies the equation for a damped harmonic oscillator with
time-dependent parameters.  
At early times, the axion mass is insignificant and
$\theta$ is approximately constant.  
When the universe cools to the critical temperature, $T_1$,
which we define via 
\begin{equation}
m_{a}(T_{1}(t_{1}))t_1\equiv 1\; ,
\label{eq:t1def}
\end{equation}
the field will begin to oscillate in its potential \cite{KT1994}.  Given 
the definition of the critical time, $t_1$, in Eq.~(\ref{eq:t1def}) \cite{Sikivie:2006ni},
\begin{equation}
t_1\simeq 2\times 10^{-7}\mathrm{s}\left(\frac{f_a}{10^{12}\mathrm{GeV}}\right)^{1/3}
\label{eq:t1}
\end{equation}
and
\begin{equation}
T_1\simeq 1 \mathrm{GeV}\left(\frac{10^{12} \mathrm{GeV}}{f_a}\right)^{1/6} \; .
\label{eq:T1}
\end{equation}
The axion field can realign only as fast as causality permits, thus the
momentum of a quantum of the axion field is
\begin{equation}
p_{a}(t_{1})\sim\frac1{t_{1}}\sim10^{-9} \textrm{eV}
\end{equation}
for $f_a \simeq 10^{12}$~GeV, which corresponds to $m_a\simeq6$~$\mu$eV, by Eq.~(\ref{eq:ma}).
Thus, this population is non-relativistic or cold.

The energy density of the scalar field around its potential minimum 
is 
\begin{equation}
\rho=\frac{f_{a}^{2}}{2}[\dot{\theta}^{2}+m_{a}^{2}(t)\theta^{2}] \; .
\label{eq:edens}
\end{equation}
By the Virial theorem,
\begin{equation}
\langle\dot{\theta}^{2}\rangle=m^{2}_a\langle\theta^{2}\rangle 
=\frac{\rho}{f_{a}^{2}} \; .
\end{equation}
As axions are non-relativistic and decoupled,
\begin{equation}
\rho\propto \frac{m_{a}(t)}{S^{3}(t)} \; .
\end{equation}
Thus, the number of
axions per comoving volume is conserved, provided the axion mass varies 
adiabatically.

The initial energy density of the coherent oscillations is
\begin{equation}
\rho_{1}=\frac12 f_{a}^{2}m^2_{a}(t_{1})\theta_{1}^{2} \; 
\label{eq:rho1}
\end{equation}
and $\theta_1$ is the initial, ``misalignment'' angle.
The energy density in axions today is
\begin{equation}
\rho_{0}=\rho_1\frac{m_{a}(t_{0})}{m_{a}(t_{1})}\frac{S^{3}(t_{1})}{S^{3}(t_{0})} \; .
\end{equation}
Using Eqs.~(\ref{eq:t1def}) and (\ref{eq:rho1}), 
\begin{equation}
\rho_0=\frac12 f_{a}^{2}\frac{m_a}{t_{1}}
\left(\frac{S(t_{1})}{S(t_{0})} 
\right)^3\theta_{1}^2\; ,
\end{equation}
which implies the
axion energy density,
\begin{equation}
\Omega_{a}\simeq 0.15\left(\frac{f_a}{10^{12} \textrm{GeV}}\right)^{\frac76}
\theta_1^2 \; ,
\label{eq:omegaa}
\end{equation}
using Eqs.~(\ref{eq:ma}), (\ref{eq:t1}) and (\ref{eq:T1}).

As the axion couplings are very small, these coherent
oscillations do not decay and make axions a good candidate for the dark matter of the
universe. 

\section{Constraints on the axion}
\label{sec:bounds}

\subsection{Laboratory bounds}
\label{ssec:labbounds}

The original PQWW axion would have been of order 100 keV mass, and possessed couplings large enough to enable the axion to have been produced and detected in conventional laboratory experiments.  These included searches for axions emitted from reactors, where axions would compete with M1 gamma transitions in radioactive decay; nuclear deexcitation experiments; beam-dump experiments; and axion decay from $1^-$ heavy quarkonia states, i.e. J/$\psi$ and $\Upsilon$.  
All results were negative, thus excluding the original PQWW axion within a decade of its prediction.  As these limits are much weaker than the current astrophysical upper bounds for both the mass and coupling to radiation, the reader is directed to earlier reviews 
\cite{Kim:1986ax,Cheng:1987gp,Rosenberg:2000wb}
, and the Particle Data Group Review of Particle Properties for discussion and annotated limits \cite{Amsler:2008zzb}
.

Searches for axions have also been performed exploiting the coherent mixing of axions with photons in experiments realized with lasers and large superconducting dipole magnets.  These include both searches for vacuum dichroism and birefringence through the production of real or virtual axions in a magnetic field \cite{Semertzidis:1990qc}
, and photon regeneration (“shining light through walls”) 
\cite{Cameron:1993mr,Robilliard:2007bq,Chou:2007zzc}.  
Axion-photon mixing will be discussed briefly further on, but the all such experiments to date have set limits on the axion-photon coupling, $\gagg\sim2 \times 10^{-7}$~GeV$^{-1}$ (weakening significantly for 
$\ma>0.5$~meV) orders of magnitude weaker than current astrophysical limits and direct searches for solar axions.  While there are no prospects for the polarization experiments to compete with these latter 
($\gagg\sim10^{-10}$~GeV$^{-1}$), a new strategy to resonantly-enhance photon regeneration may enable them to improve on these limits by up to an order of magnitude \cite{Sikivie:2007qm}
, and at least one such experiment is in preparation.

Torsion-balance techniques have enabled searches for axions through their coupling to the nucleon spin, and rigorous bounds have been set on their short-range interactions for masses below 1 meV.  While impressive tour de force experiments, how to relate these limits to expectations from the Peccei-Quinn axion is not straightforward 
\cite{Adelberger:2009zz}
.

\subsection{Cosmological bounds}
\label{ssec:cosmobounds}
Whether by the vacuum realignment mechanism, or by radiation from topological strings, the production of very light axions in the early universe implies an increasing energy density of the universe in axions for lower masses:   
$\Omega_a = \rho_a/\rho_c \propto \ma^{-7/6}$.  As the relative importance of the mechanisms is still controverted, a reliable lower bound on the axion mass should obtain where the vacuum realignment contribution becomes of order unity, $\Omega_a = \mathcal{O}(1)$.  From Eq.~(\ref{eq:omegaa}), this corresponds to $\ma\sim6 $~\meV (see section \ref{ssec:vacalign}), although there is uncertainty in this estimate itself, owing to lack of knowledge of the initial value of $\theta $ within our horizon; the estimate above based on the assumption that this parameter is of order unity.  An accidentally small value would drive the mass associated with closure density downwards, including arbitrarily small masses.  On the other hand, within the concordance model, 
$\Omega_{DM} = 0.23$, implying that the axion mass should be roughly a factor of 4 higher than cited above.  Thus the ADMX microwave cavity experiment conservatively began its search campaign at $\ma\sim2$~\meV, and continues to work upwards.  Recent discussions of the cosmological bound from vacuum realignment can be found in refs.~
\cite{Hertzberg:2008wr,Visinelli:2009zm}.

The above vacuum realignment bound applies when inflation has homogenized the
axion field over our horizon.  When the PQ scale breaks after inflation, cold 
axions are additionally produced by string and domain wall decay.  These 
extra contributions mean that we require inflation to have produced less 
axions to avoid overclosing the universe and thus, the axion mass lower bound
increases (see section~\ref{ssec:vacalign}).  As the spectrum of axions from string
radiation is debated, we review the possible bounds.  If axion strings decay 
rapidly, giving a spectrum inversely proportional to momentum, the lower bound
for the axion mass is $\sim$ 15~\meV~\cite{Hagmann:2000ja}.  If the decay is less
rapid and strings go through many oscillations, an analysis based on local strings \cite{Battye:1994au} gives a lower bound of 100~\meV.  A similar analysis based
on global strings \cite{Yamaguchi:1998gx} gives a smaller lower bound of 30~\meV.
Given that the PQ symmetry is global, it is likely that the lower 
bound of 30~\meV~is applicable if the PQ symmetry breaks after inflation and axion strings decay slowly.

\subsection{Astrophysical bounds}
In general, introducing a channel for direct or free-streaming energy loss from a star's core accelerates the star's evolution. The core will contract and heat up under the influence of gravity when axions (or other exotica) compete with the production of strongly trapped photons, whose radiation pressure acts to counterbalance gravitational pressure. Furthermore, for each stellar system, axions are excluded only over a finite range of couplings. As the axion's coupling is increased in the stellar-evolution simulations, the free-streaming lower limit of the axion's excluded couplings is reached at a point where deviations from an axion-free model first become noticeable.  However, as the coupling is further increased, the axions themselves eventually become strongly trapped; the upper limit corresponds to the regime where their influence on evolution diminishes below the threshold of observation.  A comprehensive treatment of such constraints on properties of axions and other exotica has been published by Raffelt 
\cite{Raffelt:1996wa}.  
The two most relevant astrophysical limits in framing the region of interest where axions may be the dark matter are described below.

The most stringent constraint on the axion-photon coupling presently is due to horizontal branch (HB) stars, i.e. those that are in their helium-burning phase, within globular clusters.  Globular clusters provide a cohort population of stars of all the same age; those that are seen today are have masses somewhat less than that of our sun.  By the ratio of the number observed in the HB phase to those in the red giant phase, i.e. after exhaustion of core-hydrogen burning but before the helium flash, one statistically infers an average HB lifetime.  The concordance (within 10\%) between the calculated and inferred HB lifetime precludes Primakoff production of axions ($ \gamma + Ze  \rightarrow  a + Ze $, i.e. axions produced by the interaction of a real plus virtual photon) at a level corresponding to an upper bound of  $\gagg < 10^{-10}$~GeV$^{-1}$.  A definitive study of axion cooling in stars using numerical methods aims to extend this analysis \cite{axmass}.

The lowest-lying upper bound for the axion’s mass is due to SN1987a.  Axions produced by nucleon-nucleon bremsstrahlung ($ N + N  \rightarrow N + N + a $) during the core-bounce of the protoneutron star would have competed with neutrino emission.  That the duration of the neutrino pulse observed between the IMB and Kamioka water Cherenkov detectors (19 events over 10 seconds) was in good accord with core-collapse models precludes axions in the mass range  $10^{-3}$~eV $< \ma
 < 2$~eV.  This range corresponds to the free-streaming regime; as mentioned previously, for axions above this range axions themselves are strongly trapped and thus are not effective in quenching the neutrino signal.

\section{Phase-space structure of halo dark matter}
\label{sec:phasespace}

The local velocity distribution of dark matter axions is of vital importance for direct detection.  The Axion Dark Matter eXperiment (ADMX), detailed in Section~\ref{ssec:ADMX} and ref.~\cite{CarosiNJP}, in this volume, uses a microwave cavity to search directly for axions.  The observed signal is the power output from the cavity due to axion conversion to photons, as a function of frequency.  The frequency corresponds to the energy distribution of axions undergoing conversion.  As axions are 
non-relativistic, the signal frequency is given by:
\begin{equation}
\nu(v)=\frac{m_a}{h}\left(c^2+\frac12 v^2\right) \; .
\end{equation}
The local velocity distribution thus determines the signal shape.  The signal amplitude is determined
by the density of axions of a particular energy.  Thus, the phase-space distribution determines the signal 
observed.

We expect that the dark halo of the Milky Way consists of a number of components which ADMX is capable of 
observing:
(1) a thermalized component with a Maxwell-Boltzmann velocity distribution,
(2) discrete flows, from tidal stripping of satellite halos or coherent dark matter flows crossing the halo, and
(3) overdense regions that are not gravitationally bound, known as caustics.

The ADMX search technique assumes that the rates of change of velocity, velocity dispersion
and flow density are slow compared to the time scale of the experiment.  
The ADMX medium resolution (MR) channel searches for an isothermal component of the halo as dark matter
axions.  The ADMX high resolution (HR) channel searches for signals with a narrow velocity dispersion, such as discrete flows.  

Numerical
simulations produce large halos within which hundreds of smaller clumps, or subhalos, exist \cite{Navarro:1995iw,Moore:1997sg}.
Tidal disruption of these subhalos leads to flows in the form of ``tidal tails'' or
``streams.''  The Earth may currently be in a stream of dark matter
from the Sagittarius A dwarf galaxy \cite{Freese:2003na,Freese:2003tt}.  This stream of subhalo debris 
satisfies both the requirements of small velocity dispersion and a repeatable signal and thus may be detectable by the ADMX HR channel.

Non-thermalized flows from late infall of dark matter onto the halo have
also been shown to be expected \cite{Sikivie:1992bk,Natarajan:2005ut}.  The idea behind these
flows is that dark matter that has
only recently fallen into the gravitational potential of the galaxy will have
had insufficient time to thermalize with the rest of the halo and will be present in the form of discrete flows.
There will be one flow of particles falling onto the galaxy for the first time,
one due to particles falling out of the galaxy's gravitational potential for
the first time, one from particles falling into the potential for the second
time, etc.  Furthermore, where the gradient of the particle velocity diverges,
particles ``pile up'' and form caustics.  In the limit of zero flow
velocity dispersion, caustics have infinite particle density.  The velocity
dispersion of cold axions at a time, $t$, prior to galaxy formation
is approximately $\delta v_a\sim3\times10^{-17}(10^{-5} \textrm{eV}/m_a)
(t_0/t)^{2/3}$ \cite{Sikivie:1999jv}, where $t_0$ is the present age of
the universe.  A flow
of dark matter axions will thus have a small velocity dispersion, leading to a 
large, but finite density at a caustic.  

The caustic ring model, under the assumptions of self-similarity and axial symmetry,
predicts that the Earth is located near a caustic
feature \cite{Sikivie:2001fg}.  This model, fitted to rises in the Milky
Way rotation curve and a triangular feature seen in the IRAS maps, predicts 
that the flows falling in and out of the halo for the fifth time
contain a significant fraction of the local halo density.  The
predicted densities are $1.5\times10^{-24}$ g/cm$^3$ and 
$1.5\times10^{-25}$ g/cm$^3$ \cite{Duffy:2008dk}, comparable to the local 
dark matter density of $9.2\times10^{-25}$ g/cm$^3$ predicted by Gates
et al.\ 
\cite{Gates:1995dw}.  The flow of the greatest density is known
as the ``Big Flow.''  

A general treatment of the phase-space structure of dark matter halos, 
which does not require assumptions of self-similarity or symmetry, has recently been developed 
\cite{Afshordi:2008mx}.  This treatment studies the statistics of dark matter caustics in the tidal debris remaining from mergers
of smaller halos to form galaxies and from the primordial coldness of dark matter.  While more general than the approach of ref.~\cite{Duffy:2008dk}, this
treatment only results in a statistical distribution and does not give specific predictions for our galactic halo.

Additionally, numerical methods have been developed to study caustics and flows in dark matter
halos.  To date, most numerical simulations are too coarse-grained to resolve
caustic structure,
although its presence can be observed when special techniques are used 
\cite{DoroshkevichMNRAS,KlypinMNRAS,CentrellaNature,MelottNature,Stiff:2003tx}.  
The recent work of ref.~\cite{Vogelsberger:2007ny} 
predicts at least $10^5$ discrete streams near our Sun, although specific predictions are not made for the
stream densities and velocities.

It has also recently been shown that dark matter axions can exist in the form of a 
Bose-Einstein condensate (BEC)\cite{Sikivie:2009qn}.  If this is the case, the formation of caustics is suppressed 
within the BEC.  However, vortices are expected to form at the center of galactic halos, due to their net rotation.  
Within a vortex, axion dark matter will exist in the normal phase and flows and caustics will still be present.

The possible existence of discrete flows
provides an opportunity to increase the discovery potential of ADMX.  A 
discrete axion flow produces a narrow peak in the spectrum of microwave
photons in the experiment and such a peak can be searched for with higher
signal-to-noise than the signal from axions in an isothermal halo.
If such a signal is found, it will provide detailed information on
the structure of the Milky Way halo.

\section{Experimental searches for dark-matter axions}
\label{sec:exp}

\subsection{Axion-photon mixing}

As pseudoscalars, axions can be produced by the interaction of two photons, one of which can be virtual, $\gamma + \gamma* \rightarrow a$ , the process being known as the Primakoff effect \cite{Primakoff}.  This implies that photons and axions may mix in the presence of an external electromagnetic field, through the Lagrangian density
of Eq.~(\ref{eq:aEB}).  
In all axion searches based on the Primakoff effect to date, $\mathbf{E}$ represents the electric field of the real photon, and $\mathbf{B}$ is an external magnetic field.   Although the opposite combination is possible, it is vastly easier to produce and support a static magnetic field than the equivalent electric field; a 1 T magnetic field being equal to an electric field of 30 MV/cm in gaussian units.  In fact, fields of order 10 T are readily achieved today with superconducting magnets.   The coherent mixing of axions and photons within magnetic fields of large spatial extent enables searches of exceedingly high sensitivity, although there is yet no experimental strategy capable of reaching the standard Peccei-Quinn axion over the entire open range.

A general formulism for axion-photon mixing in external magnetic fields, including plasma effects, is found in Ref.~\cite{Raffelt:1987im}.

\subsection{The microwave cavity experiment for dark-matter axions}
\label{ssec:ADMX}

In 1983, Sikivie proposed two independent schemes to detect the axion based on the Primakoff effect \cite{Sikivie:1983ip,Sikivie:1985yu}.  The first was a search for axions constituting halo dark matter by their resonant conversion to RF photons in a microwave cavity permeated by a strong magnetic field.  Tuning the cavity to fulfill the resonant condition, $h\nu = m_ac^2(1 + \mathcal{O}(\beta^2\sim10^{-6}))$, and assuming axions saturate the galactic halo, the conversion power from an optimized experiment is given by:
\begin{equation}
\label{eqn-convpower}
P=g_{a\gamma\gamma}^{2}
\frac{VB^{2}\rho_{a}Q}{m_{a}}\mathrm{,}
\end{equation}
where $B$ is the strength of the magnetic field, $V$ the cavity volume and $Q$ is the cavity quality 
factor.  The most sensitive microwave cavity experiment (and in fact the only one currently in operation) is ADMX at Lawrence Livermore National Laboratory. This search has excluded axions of KSVZ axion-photon coupling as the local halo dark matter, for a narrow range of masses $1.9 < m_a < 3.4$~$\mu$eV.   

The anticipated conversion power is miniscule, even for the largest 
superconducting magnets feasible; for ADMX the signal expected is of 
order $10^{-22}$ watts.  Furthermore, as the experiment will necessitate tuning orders of magnitude of frequency in small frequency steps, there are limits on how long one may integrate at each frequency to improve the signal-to-noise, as governed by the Dicke radiometer equation \cite{radiometer}:
\begin{equation}
\frac{s}{n}\equiv\frac{P_s}{P_n}=\frac{P_s\sqrt{\Delta\nu t}}{kT_n}\;.
\end{equation}
Here  $s/n$  is the signal-to-noise ratio,  $\Delta\nu$ the bandwidth of the signal,  $t$  the integration time, and   $P_s$  and  $P_n$  the signal and noise power respectively. This puts a clear premium on reducing the total system noise temperature, which is the sum of the physical temperature and the equivalent electronic noise temperature of the amplifier,   $T_n  =  T_{phys} +  T_{elec}$.   ADMX has recently completed an upgrade from conventional heterojunction field-effect transistors (HFETs, or HEMTs) with an noise equivalent temperature of $T_{elec} \sim 2$~K, to SQUID amplifiers, whose noise equivalent noise temperatures can reach the quantum limit, $T_{elec} \sim 50$~mK at 750 MHz  when cooled to comparable physical temperatures.  This strategy will enable them ultimately to reach the DFSZ model axions, as well as cover the open mass range much faster.  The microwave cavity experiments is described by Carosi elsewhere in this volume 
\cite{CarosiNJP}.

\subsection{Other current axion searches}

In the same report, Sikivie also outlined how to detect axions free-streaming from the Sun's nuclear burning core \cite{Sikivie:1983ip,Sikivie:1985yu}.  Axion production would be dominated by the Primakoff process  $\gamma + Ze 
\rightarrow a + Ze$; for KSVZ axions, the integrated solar flux at the Earth would be given by $F_a =  7.4 \times 10^{11} m_a^2$[eV]~cm$^{-2}$s$^{-1}$, emitted with a thermal spectrum of mean energy $\sim 4.2$~keV.  For relativistic axions, the conversion probability to photons of the same energy in a uniform magnetic field is given by $P(a \rightarrow \gamma ) = \Pi = (1/4)(g_{a\gamma\gamma}BL)^2 |F (q)|^2$,  where $B$ is the strength of the magnetic field,  and $L$ its length\footnote{A useful mnemonic in rate estimates for experiments is that, within a few percent, the factor $(g_{10}B_{10}L_{10})^2 \approx 10^{-16}$, where $g_{10} \equiv 10^{-10}$~GeV$^{-1}$, $B_{10} \equiv 10$~T, and $L_{10} \equiv 10$~m.}.  $F(q) \equiv \int dx e^{–iqx} B(x) /B_0L$ represents the form-factor of the magnetic field with respect to the momentum mismatch between the massive axion and massless photon of the same energy,  $q = k_a - k_{\gamma} = (\omega^2-m_a^2)^{1/2}-\omega \sim m_a^2/2\omega$.  $|F(q)|$ is unity in the limit $ql << 2\pi$, but oscillates and falls off rapidly for $ql > 2\pi$, where the axions are no longer sufficiently relativistic to stay in phase with the photon for maximum mixing.

Utilizing a prototype LHC dipole magnet as the basis for its ‘axion helioscope’, the CAST collaboration have recently published the best limits on the solar axions, $\gagg < 0.88 \times 10^{-10}$~GeV$^{-1}$, valid for $\ma <  10^{-2}$~eV \cite{Andriamonje:2007ew}, slightly more stringent than those derived from horizontal branch stars.  This collaboration has also pushed the sensitivity of the search upward in mass into the region of axion models, by introducing a gas ($^4$He) of variable pressure into the magnet bore.  In this case, the plasma frequency 
$\omega_p = (4\pi\alpha N_e/m_e)^{1/2} \equiv m_\gamma$ endows the x-ray photon with an effective mass; thus full coherence of the axion and photon states can be restored, and the theoretical maximum conversion probability achieved for any axion mass, by the filling the magnet with a gas of the appropriate density \cite{vanBibber:1988ge}.  The mass range can thereby be extended upwards in scanning mode, by tuning the gas pressure in small steps to as high as feasible.  In this manner, axions have now been excluded part-way into the Peccei-Quinn model band, $\gagg < 2.2 \times 10^{-10}$~GeV$^{-1}$ (95\% c.l.), valid for $\ma <  0.4$~eV \cite{Arik:2008mq}.  This phase of the experiment continues with $^3$He gas which will permit probing of yet higher masses; for further details, see Zioutas elsewhere in this volume \cite{ZioutasNJP}.  

Purely laboratory bounds on axions or generalized pseudoscalars have also been established without relying on either astrophysical or cosmological sources. In photon regeneration (``shining light through walls") axions are coherently produced by shining a laser beam through a transverse dipole magnet, and reconverted to real photons in a colinear dipole magnet on the other side of an optical barrier \cite{VanBibber:1987rq,Anselm:1986gz}.  The probability to detect a photon per laser photon is given by $P(a \rightarrow \gamma \rightarrow a ) = \Pi^2$.  
While current limits from photon regeneration ($\gagg <  2 \times 10^{-7}$~GeV$^{-1}$, $\ma < 0.5 \times 10^{-3}$~eV) \cite{Cameron:1993mr,Robilliard:2007bq,Chou:2007zzc} have not so far been competitive with solar searches, the scheme may be resonantly enhanced utilizing actively locked Fabry-Perot optical cavities to strengthen limits potentially by an order of magnitude beyond those of CAST and Horizontal Branch stars \cite{Sikivie:2007qm}.

\begin{figure}
\begin{center}
\includegraphics[height=0.5\textheight]{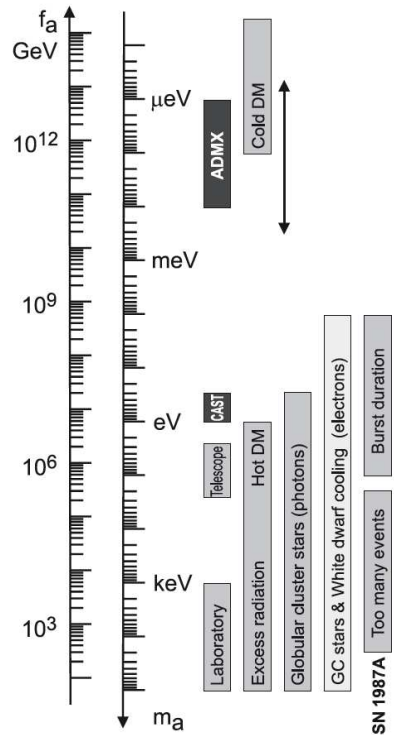}
\end{center}
\caption{Constraints on the PQ-scale $f_a$ and corresponding $\ma$ from astrophysics, cosmology and laboratory experiments.  The light grey regions are most model-dependent.  From ref. \cite{Amsler:2008zzb}.}
\label{fig:1}
\end{figure}

\section{Summary}
\label{sec:sum}

The current state of the standard QCD axion is shown in Figure~\ref{fig:1} \cite{Amsler:2008zzb}.  While it represents an oversimplification of the situation, insofar as the various experimental and observational limits on mass are model-dependent, the central point is that there is a substantial window for axionic dark matter, and that the upgraded ADMX will be able to cover about two of the three decades in mass.  One should be open to surprise from experiments such as CAST which are looking in regions of mass and coupling constant where axions are not expected to be the dark matter, but could find axion-like pseudoscalars associated with our first forays into beyond-Standard Model phenomenology.  

\section*{Acknowledgments}

This work at Lawrence Livermore National Laboratory was supported in part by the U.S. Department of Energy under Contract No. DE-AC52-07NA27344 and is approved for publication under LLNL-JRNL-412070. At Los Alamos National Laboratory, this work was supported in part by the
National Nuclear Security Administration of the U.S.
Department of Energy under Contract No. DE-AC52-06NA25396 and is approved for publication under LA-UR 09-02007.  The support of the Laboratory Directed Research and Development Program for enabling technology development at both Lawrence Livermore and Los Alamos National Laboratories is gratefully acknowledged. 

\section*{References}

\bibliographystyle{unsrt}
\bibliography{References}

\end{document}